\documentclass[aps,prb,superscriptaddress,onecolumn]{revtex4-2}

\usepackage{hyperref}       
\usepackage{url}            
\usepackage{amsmath,amssymb}
\usepackage{booktabs}       
\usepackage{amsfonts}       
\usepackage{graphicx}
\usepackage{tikz}
\usepackage[caption=false]{subfig}
\usepackage{siunitx}
\usepackage{ulem} 

\usepackage[export]{adjustbox}
\usepackage{xcolor}
\graphicspath{ {./fig/} } 

\usepackage{booktabs}

\usepackage{crop}%
\usepackage{amsmath,amssymb,amsfonts,upref} 
\usepackage{amsthm}
\usepackage{marginnote}
\usepackage{soul}
\RequirePackage{graphicx}
\usepackage[figuresright]{rotating}
\usepackage{color}
\usepackage{colortbl}

\newcommand{\ket}[1]{|#1\rangle}

\begin{document}

\title{Experimental demonstration of robustness under scaling errors for superadiabatic population transfer in a superconducting circuit}

\author{Shruti Dogra}
\affiliation{Department of Applied Physics,
	Aalto University School of Science, P.O. Box 15100, FI-00076 AALTO, Finland}

\author{Antti Veps\"al\"ainen}
\affiliation{Department of Applied Physics,
	Aalto University School of Science, P.O. Box 15100, FI-00076 AALTO, Finland}

\author{Gheorghe Sorin Paraoanu}
\email{sorin.paraoanu@aalto.fi}
\affiliation{Department of Applied Physics,
	Aalto University School of Science, P.O. Box 15100, FI-00076 AALTO, Finland}




\begin{abstract} We study experimentally and theoretically the transfer of population between the ground state and the second excited state in a transmon circuit by the use of superadiabatic stimulated Raman adiabatic passage (saSTIRAP). We show that the transfer is remarkably resilient against variations in the amplitudes of the pulses (scaling errors), thus demostrating that the superadiabatic process inherits certain robustness features from the adiabatic one. In particular, we put in evidence a new plateau that appears at high values of the counterdiabatic pulse strength, which goes beyond the usual framework of saSTIRAP.
\end{abstract}
\maketitle


\section{Introduction}

In recent years, shortcuts to adiabaticity have emerged as a powerful technique of quantum control. While in the past both adiabatic and Rabi types of control have been of tremendous utility, modern quantum control based on shortcuts to adiabaticity aims at combining the advantageous features of both of these techniques. Specifically, adiabatic processes are known to be insensitive to small errors in the shape of the pulses, but their time of operation is slow, limited by the adiabatic theorem. On the other hand, control methods based on Rabi oscillations, as typically used for quantum gates, are fast but needs precise timings and pulse-shape control. Thus, processes that are both fast and resilient against variations of the pulse parameters are of significant interest for the field of quantum information processing. Here we discuss one such process, the superadiabatic correction of STIRAP (stimulated Raman adibatic passage) and observe its robustness features in an experiment of population transfer in a transmon qubit.


\maketitle


Adiabatic control methods based on the stimulated Raman adiabatic passage (STIRAP) have been widely studied and integrated into various experimental schemes~\cite{Bergmann_2019,shore-review-2017}. Recently, a lot of interest has been raised by the concept of superadiabatic or transitionless process, introduced by Demirplak and Rice~\cite{Demirplak03, Demirplak05} and by Berry~\cite{Berry09}. Superadiabatic processes belong to  a more general class of protocols generically referred to as shortcuts to adiabaticity. 
When applied to STIRAP, this results in the so-called saSTIRAP protocol, where a counteradiabatic pulse is applied in addition to the usual pump and Stokes STIRAP pulses~\cite{arimondo_sa_stirap, antti-2019}.

Due to the fact that experiments are subjected to various sources of errors and imperfections, a key task is to characterize the robustness of these protocols. For any practical implementation, we would need to understand if the generic features of a process typically designed for ideal operation remain valid in realistic conditions. For STIRAP, the effects of noise have been studied already theoretically as well as on some experimental platforms. For example, in superconducting circuits the reduction of population transfer due to broadband colored noise was discussed in Ref. \cite{Falci2013,Falci2016}.  
In Ref.~\cite{kostas-pra-2020}, it was shown that saSTIRAP works well in the noisy environment consisting of dissipation and Ornstein-Uhlenbeck dephasing. It was found that the performance of shortcuts to adiabaticity decreases with the correlation time of the  Ornstein-Uhlenbeck noise \cite{Paspalakis2020}, see also \cite{Pope_2019}. In the context of shortcuts to adiabaticity using the Lewis-Riesenfeld method, various noises (white noise, Ornstein-Uhlembeck noise, flicker noise, constant error) in the spring constant of the trap have been studied in Ref. \cite{Muga2011,Lu2014}. 
Other errors, such as the accordion noise in the wave vector of the trap, and errors in the amplitude and phase of the trap have been addressed systematically in Ref. \cite{Lu2020}. 
Another approach is to mitigate decoherence by incorporating it into the adiabatic evolution and into the STA protocols~\cite{Paternostro-njp-2014, Paternostro-scirep-2014} and to speed up the dynamics by considering non-Hermitian Hamiltonians~\cite{Chen-pra-2016, Chen2018}.

For superadiabatic processes, which is our main interest in this work, several theoretical studies have been performed. Already in 2005, Demirplak and Rice have addressed the sensitivity of the counterdiabatic technique to errors such as
pulse  width, location  of  the  peak,  and  intensity, demonstrating that a window of errors exist where high-fidelity transfer can be maintained \cite{Rice2005}. In the case of two-level systems, the robustness of several driving schemes against off-resonance effects, as well as the tradeoff with respect to the speed of the process, has been analyzed recently in Ref.~\cite{Alsing2020}. The effect of dephasing for superadiabatic STIRAP in three-level system has been considered in \cite{Issoufa2014}. In Ref. \cite{Yu2018} transfer protocols in a system of coupled 1/2 spins have been analyzed, where the systematic errors due to shifts in the magnetic field and due to the Dzyalozynskii-Moryia interaction between the two spins are treated perturbatively.

In the present work we study experimentally and theoretically the robustness of sa-STIRAP in a three-level system with respect to scaling errors. 
For a three-level system, the superadiabatic STIRAP (saSTIRAP) is characterized by the simultaneous application of three pulses, the pump and Stokes pulses of STIRAP (coupling the levels $|0\rangle-|1\rangle$ and $|1\rangle-|2\rangle$ respectively) and the counterdiabatic pulse coupling into  the  $|0\rangle-|2\rangle$ transition. These three pulse drive the three-level transmon concurrently in a loop (also called $\Delta$-driving). Scaling errors are errors which come into play due to technical limitations, causing miscalibrations during the experimental implementation. For a given pulse shape (say a Gaussian), the errors can originate from the miscalibration of the pulse-amplitude, pulse-width, phase and relative placement of the pulse in the pulse sequence.  Robustness under scaling has also been discussed theoretically in Ref. \cite{SorensenMolmer2018}, where a cost functional has been introduced
while observing random scaling with STIRAP pulses, 
and in Ref.~\cite{Barnum-2020}, where optimal STIRAP population transfer 
was observed as a function of the relative delay between the pump and Stokes pulses.

Here we analyse the robustness of saSTIRAP in wide ranges of experimental parameters, providing a systematic characterization of robustness under scaling errors. We also analyse the imperfections related to the counterdiabatic term in detail. We first present a method of characterizing this term based on the Wigner-Ville spectrum. Then we focus on the phase of the drive, and the saSTIRAP performance resulting from the use of two-photon resonant drive as compared to the direct coupling between the initial and the target states.
We characterize the experimentally-achieved quantum speed limit during the state evolution in STIRAP and saSTIRAP based on the Bures distance. Next, we observe the formation of plateaus of nearly-constant populations in the final state as a function of STIRAP and counterdiabatic pulse areas. This demonstrates that saSTIRAP is largely insensitive to scaling errors. Furthermore, we extend our study to counterdiabatic pulsea areas significantly larger than the superadiabatic value of $\pi$. Suprisingly, we observe the existence of regions where population transfer still occurs with high efficiency.


\section{Experimental setup}

\begin{figure}[ht]
	\includegraphics[width = 0.7\columnwidth]{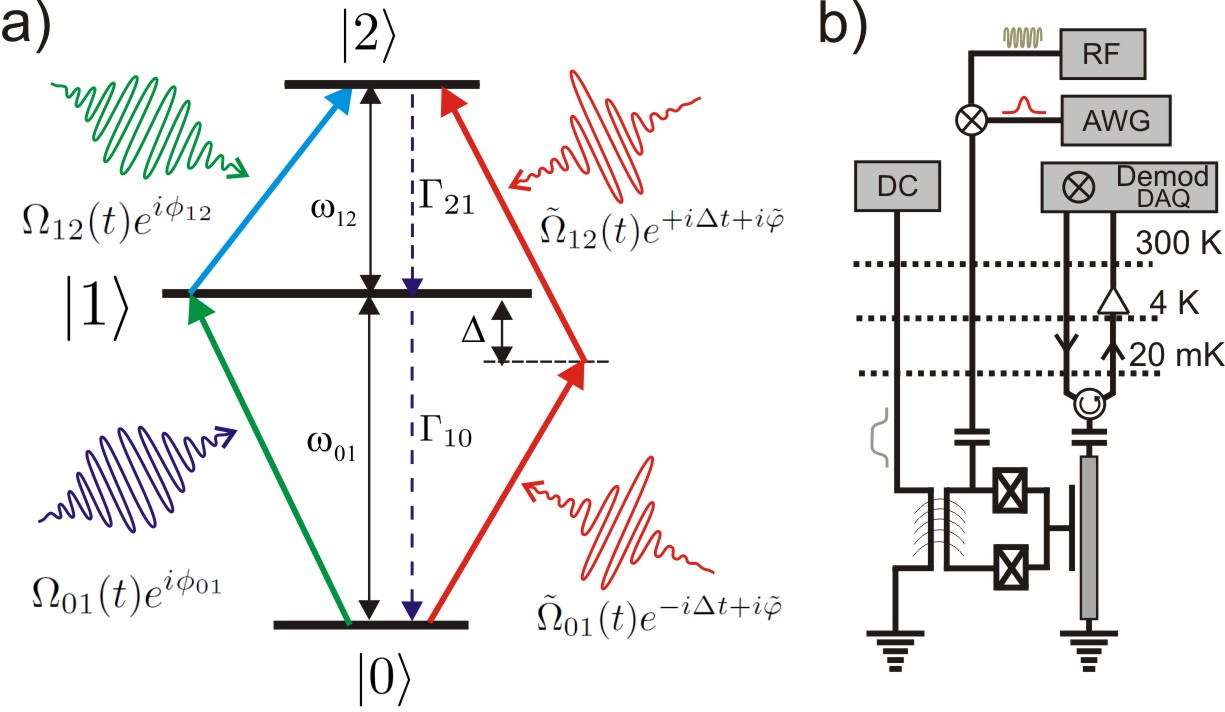}
	\caption{a) The three-level system is driven by two microwave fields applied resonantly to the 0-1 and 1-2 transitions and an additional field applied off-resonantly with respect to state $|1\rangle$. 
		b) Simplified schematic of experiment, showing the transmon device coupled to a readout resonator and placed at the mixing chamber in a dilution refrigerator, as well as the electronic blocks used to control the system. } \label{Fig-exp_setup}
\end{figure}

The experiments are performed on the three lowest energy levels ($|0\rangle$, $|1\rangle$, $|2\rangle$) of an artificial atom (see Fig. \ref{Fig-exp_setup}) constituted by a pair of Josephson junctions with transition frequencies $\omega_{01}=2\pi \times 7.395$ GHz and $\omega_{12}=2\pi \times 7.099$ GHz, with anharmonicity $2\Delta = (\omega_{01}-\omega_{12})= 2\pi \times 296$ MHz.
Here, we consider Gaussian shaped pulsed fields: $\Omega_{01}(t)=\Omega_{01}e^{-t^2/2\sigma^2}$ and $\Omega_{12}(t)=\Omega_{12}e^{-(t-t_s)^2/2\sigma^2}$, which couple with transitions $|0\rangle-|1\rangle$ and $|1\rangle-|2\rangle$ respectively. These Gaussians have the same values of standard deviation $\sigma$ and are being truncated at $\pm n\sigma$. The counter-intuitive STIRAP pulse sequence requires the drive $\Omega_{12}(t)$ to set in before the drive $\Omega_{01}(t)$ with an overlap for time, $t_s=-\kappa \sigma$. The counterdiabatic drive $\Omega_{02}(t)=2\dot{\Theta}$, where $\Theta=\tan^{-1}(\Omega_{01}(t)/\Omega_{12}(t))$~\cite{stirap_ours}. In two-photon resonance excitation, amplitudes of the effective couplings $\tilde{\Omega}_{12}=\sqrt{2} \tilde{\Omega}_{01}$ and phases of the drives, $\Phi=\phi_{01}+\phi_{12}+\phi_{20}=-\pi/2$.

In the rotating wave approximation applied to the transmon circuit,
\begin{eqnarray}
	H_{01}(t) &=& \frac{\hbar}{2}\Omega_{01}(t) e^{i \phi_{01}} |0\rangle \langle 1| + h.c. , \label{eq:H01}\\
	H_{12} (t) &=& \frac{\hbar}{2}\Omega_{12}(t) e^{i \phi_{12}} |1\rangle \langle 2| + h.c.,  \label{eq:H12}
\end{eqnarray}

The 0-2 coupling is
\begin{eqnarray}
	\tilde{H}_{\rm 2ph}(t) &=& \frac{\hbar}{2}\left[\tilde{\Omega}_{01}(t)e^{-i\Delta t + i\tilde{\varphi}}|0\rangle \langle 1| + \tilde{\Omega}_{12}(t)e^{+i\Delta t + i \tilde{\varphi}}|1\rangle \langle 2 |\right] + h.c.
	\label{eq:H2ph}
\end{eqnarray}
resulting in a 0-2  Rabi frequency $\Omega_{02} = \tilde{\Omega}_{01}\tilde{\Omega}_{12}/(2\Delta)$ and 
phase $\phi_{02} = 2\tilde{\varphi}  + \pi$,
\begin{eqnarray}
	H_{02}(t) &=& -\frac{\hbar \tilde{\Omega}_{01}\tilde{\Omega}_{12}}{4 \Delta} e^{2i\tilde{\varphi}}|0\rangle \langle 2| + h.c.
	\label{eq:H02}
\end{eqnarray}

The evolution of the system can be formulated as a dynamical map with trace-class
	generator $\mathbb{L}$,
	\begin{equation}
	\dot{\rho}(t) = \mathbb{L}[\rho(t)], \label{eq:2.5}
	\end{equation}
	where $\mathbb{L}[\rho (t)] = (i/\hbar )[\rho (t) , H(t)] + \mathcal{L}[\rho (t)]$, with $\cal{L}$ the Lindblad superoperator,
	\begin{equation}
	{\cal L}[\rho (t) ] = \sum_{{\rm c}}\Gamma_{{\rm c}}\left( \Lambda_{{\rm c}}\rho (t)\Lambda_{{\rm c}}^{\dag} - \frac{1}{2}\{\Lambda_{{\rm c}}^{\dag}
	\Lambda_{{\rm c}}, \rho (t) \}\right),   \label{eq:2.6}
	\end{equation} 
	where ${\rm c}$ is an index denoting the dissipation channel with Lindblad operator $\Lambda_{c}$.
	For this sample, the dissipation results predominantly from relaxation processes, namely two decay channels are present: the decay from level 2 to level 1 with relaxation rate $\Gamma_{21}$ and $\Lambda_{21}=|2\rangle \langle 1|$ and the decay rate from level 1 to the ground state, with 
	$\Lambda_{10}=|1\rangle \langle 0|$.
	We get $\Gamma_{10} =$ 5.0 MHz and $\Gamma_{21} =$ 7.0 MHz, with dephasing times being dominated by the energy relaxation for this sample. For a three-level system \cite{EIT_abdumalikov,ourPRB}, the Lindbladian is ${\cal L} [\rho ] = -\Gamma_{21} \rho_{22} |2\rangle\langle 2| - (\Gamma_{10} \rho_{11} - \Gamma_{21} \rho_{22}) |1\rangle\langle 1| + \Gamma_{10} \rho_{11} |0\rangle\langle 0|$.

\section{Numerical analysis and optimization of drive parameters}

The superadiabatic STIRAP is driven jointly by the drives $H_{01}(t)$, $H_{12}(t)$, and $\tilde{H}_{\rm 2ph}(t)$. Thus, a precise saSTIRAP implementation involves a balance between the choice of several parameters. In the following subsections, we analyse the robustness of saSTIRAP with respect to each of these parameters with wide ranges of experimentally feasible values. It is noteworthy that saSTIRAP is found to display remarkable resilience against variation in these parameter values.
 We perform a quantitative analysis of the saSTIRAP protocol using the population transferred to the second excited state $p_2$ as the figure of merit.

\subsection{Wigner-Ville analysis of the counterdiabatic pulse}

The Wigner-Ville (WV) spectrum is a mathematical tool of great utility for characterizing transient processes, allowing the analysis of non-stationary signals in both time and frequency domains~\cite{cohen1989}. This allows us to identify all frequency components contained in the autocorrelation function at any time. Note that for non-stationary noises the usual tools are generally not applicable, as the Wiener-Kinchin theorem, which connects the correlation function with the power spectrum, is not valid. Here we employ the Wigner-Ville analysis for obtaining the spectral content of the correlations of the counterdiabatic pulse. The reason for this choice is that the counterdiabatic pulse is the key component of our scheme and at the same time the pulse most difficult to generate precisely.

For a time-dependent complex variable $x(t)$ with  correlation function $\langle x(t + \tau /2 )x^{*}(t - \tau /2 )\rangle $ we define the Wigner-Ville spectrum 
\begin{equation}
	S_{WV} (\omega, t) = \int_{-\infty}^{\infty} e^{- i \omega \tau}  \langle x(t + \tau /2 )x^{*}(t - \tau /2 )\rangle d\tau .
\end{equation} 

For stationary systems, one recovers the usual definition of power spectral density as $S_{xx}(\omega ) = S_{WV} (\omega, 0)$. The instantaneous power at any moment $t$ can be obtained by summing the Wigner-Ville spectrum over all infinitesimal bandwidths 
\begin{equation}
	\langle |x(t)|^{2} \rangle = \frac{1}{2\pi} \int_{-\infty}^{\infty} d \omega S_{WV}(\omega , t).
\end{equation}

\begin{figure}[ht]
	 	\begin{center}
	 		\includegraphics[width = 0.8\columnwidth]{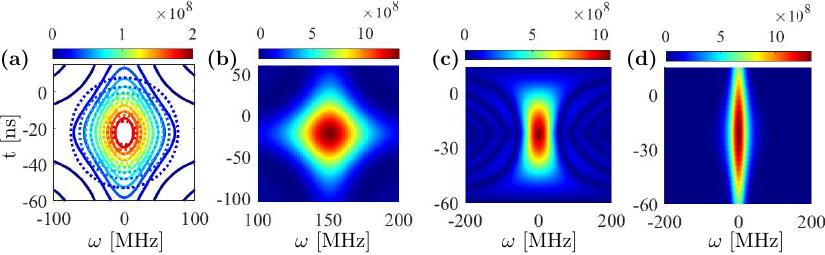} 
	 	\end{center}
	 	\caption{(a) The contour plot of the Wigner-Ville spectrum Eqs.~(\ref{eq-WV}) as a function of frequency and time is shown with continuous lines for $\kappa=-1.5$, $\sigma=30$ ns. The doted lines represent the approximation described in the text. (b) Surface maps of the WV spectrum obtained from self-correlations of the 
	 		drive $\tilde{\Omega}_{12}(t) e^{i\Delta t + i \tilde{\varphi}}$ with $\tilde{\varphi}=-\pi/4$.
	 		The WV spectrum of $\tilde{\Omega}_{01}(t)$ with truncated saSTIRAP sequence at $n=2$ is presented in panel (c) and the corresponding WV spectrum of $\tilde{\Omega}_{01}(t)$ with infinite extent is shown in panel (d).}
	 	\label{fig-WV}
	 \end{figure}

In the case of saSTIRAP, all ac-Stark energy shifts are related to the same function, the two-photon coupling. Thus, all cross-correlations between noises become self-correlations of the effective 0 - 2 coupling
\begin{equation}
	\Omega_{02} (t) = -\frac{t_{s}}{\sigma^{2}} \frac{1}{\cosh \left[-\frac{t_{s}}{\sigma^{2}}\left( t - \frac{t_{s}}{2} \right) \right]}.
	\label{eq:omega02t}
\end{equation}
Thus we can obtain the Wigner-Ville spectrum 
	\begin{eqnarray}
		S_{WV}^{(02)} (\omega, t) &=& \int_{-\infty}^{\infty} e^{- i \omega \tau}  \langle \Omega_{02}(t + \tau /2 )\Omega_{02}(t - \tau /2 )\rangle d\tau, \nonumber \\
		&=& \frac{2|ts|}{\sigma^2} \int_{-\infty}^{\infty} \frac{ e^{-i \frac{\sigma^2}{|t_s|} \omega \tau}}{\cosh[\frac{2t_s}{\sigma^2}(t-\frac{t_s}{2})] + \cosh[\frac{t_s}{\sigma^2}\tau]} d\tau.
		\label{eq-WV}
	\end{eqnarray}
	This function is quantified numerically and plotted versus frequency $\omega$ and time $t$ as shown by continuous lines in Fig.~\ref{fig-WV}(a). An approximation of the Wigner-Ville distribution function for small $t$ and $\tau$ i.e. with vanishing self-correlations at larger $\tau$, is $S_{WV}^{(02)} (\omega, t) \approx 2\sqrt{\pi}|t_{s}|/\sigma^2 e^{- \frac{t_{s}^{2}}{\sigma^{4}}(t-\frac{t_{s}}{2})^{2}} e^{-\frac{\sigma^{4}}{t_{s}^2}\omega^{2}}$.
	 This is plotted as a function of frequency and time with dotted lines in Fig.~\ref{fig-WV}(a).
	 
	 To get an understanding of the behaviour of this spectrum, we noticed that it has maximal values at $\omega=0$ and $t=t_s/2$, otherwise $S_{WV}^{(02)} (\omega, t)$ drops fast to zero. The WV spectrum in the time-domain for $\omega=0$ is given by
	 \begin{eqnarray}
	 	S_{WV}^{(02)} (0, t) &=& \frac{8 t_s^2}{\sigma^4} \left  (t-\frac{t_s}{2} \right) \, \textrm{cosech}\left[\frac{2|t_s|}{\sigma^2}\left(t-\frac{t_s}{2}\right)\right],
	 	\label{eq:wv0freq}
	 \end{eqnarray}
	 and in the frequency domain corresponding to $t=t_s/2$ it is given by
	 \begin{equation}
	 	S_{WV}^{(02)} (\omega, t_s/2) = 4\pi \omega \, \textrm{cosech}[\sigma^2 \pi \omega/|t_s|].
	 \end{equation}

	 Each of these attain a maximum value of $4|t_s|/\sigma^2$, which is four times the maximum of $\Omega_{02}(t)$ drive.
	 The zero frequency WV $S_{WV}^{(02)}(0,t)$ is narrower than $\Omega_{02}(t)$, which in terms of full width at half maxima (FWHM) reads FWHM[$S_{WV}^{(02)}(0,t)$]$<$FWHM[$\Omega_{02}(t)$]. 
	 This feature appears in the frequency domain as well, where  $S_{WV}^{(02)}(\omega,t_s/2)$ at $t=t_s/2$ is narrower than the Fourier transform of the drive, $FT[\Omega_{02}(t)]= \Omega_{02}(\omega) \equiv \sqrt{\pi/2}/\cosh(\pi \sigma^2 \omega/(2|t_s|))$.

	 The experimental implementation of the $\Omega_{02}(t)$ drive has been done via two-photon resonant drive as per Eq.~(\ref{eq:H2ph}) with  $\tilde{\Omega}_{01}(t) = \tilde{\Omega}_{12}(t)/\sqrt{2} = \sqrt{\sqrt{2}\Delta \Omega_{02}(t)}$. These two simultaneous pulses are off-resonant from the $|0 \rangle-|1 \rangle$ and $|1\rangle-|2\rangle$ transitions by $-\Delta$ and $\Delta$ respectively, as shown in Fig.~\ref{Fig-exp_setup}(a). The WVD of the two-photon resonance is linearly shifted in frequency due to time-dependent phase factor $e^{\pm i \Delta t}$, such that its maximum lies at $(t=t_s/2, \omega=\pm \Delta)$ as shown in Fig.~\ref{fig-WV}(b) for  $\tilde{\Omega}_{12}(t)$.

	 Two-photon resonant pulses have $1/\sqrt{\cosh}$ pulse envelopes, which are generated using an arbitrary waveform generator with a sampling rate of 1 G-samples/s. The simulations of the WVD of this smooth function shown here are performed with discrete number of time points $~100-200$, which is in the same range as the number of experimental pulse points. In the ideal situation, as shown in Fig.~\ref{fig-WV}(b,d), no spurious excitations are observed. However, in the actual implementation, the pulses are truncated. Thus, in this case the integration limits of the auto-correlation function narrows down from $(-\infty, +\infty)$ to $(t_i+\tau/2, t_f-\tau/2)$ as $x(t\pm\tau/2)$ vanishes for $t_f < (t\pm \tau/2) < t_i$. 
	 This truncation of the drives leads to a ripple effect~\cite{Narasimhan1998}, which can be troublesome in frequency domain causing spurious excitations as shown in Fig.~\ref{fig-WV}(c) for n=2. Unlike the Fourier transform, the WV spectrum provides the complete picture of these ripples in both time and frequency domains. It is noteworthy that the truncation has more adverse effects close to the beginning and the end of the drive. Also, corresponding to different time points of the drive, these ripples shift along the frequency axis as shown in Fig.~\ref{fig-WV}(c). Thus abrupt truncation not just leads to a broader bandwidth but also causes spurious excitations for a varied range of frequencies. We avoid the generation of these ripples by using an optimal truncation of the drives.

	 An approximation of the spectral power density in the frequency domain, $\int_{t_i}^{t_f} S_{WV}(\omega, t) dt$ gives an average power distribution among various frequencies. An average of this quantity over frequency provides the average power associated with the drive. Using this definition of average power, we find that the truncation of our pulse sequence for $n\geq 4$ does not change significantly this average power. 
	 More importantly, the ripples in Fig.~\ref{fig-WV}(c) are small for $n=3$ and almost vanish for $n\geq4$. Thus, when evaluating the experimental feasibility, we conclude that a truncation at $n=3$ is a reasonable compromise, achieving both short total transfer time and avoiding spurious excitations.


\subsection{Optimal phases of the drives}

\begin{figure}[htb]
	\centering
	\includegraphics[scale = 1.2]{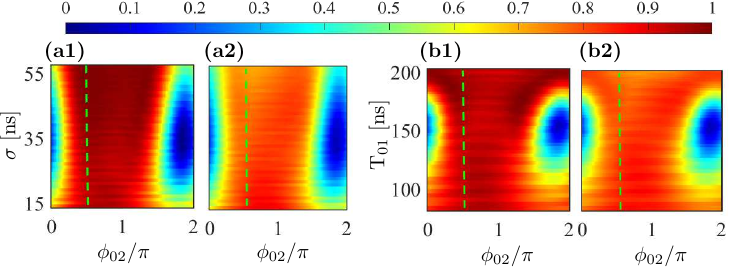} 
	\caption{Maximum attained population $p_2$ is simulated for saSTIRAP as a  
			function of (a) initial phase of the counter diabatic drive ($\phi_{02}$) and standard deviation of the Gaussian drives ($\sigma$)
			with $\kappa=2.7$, $\phi_{01}=\phi_{12}=0$.
			(b) initial phase of the counter diabatic drive ($\phi_{02}$) and total time taken by the saSTIRAP drives (T$_{01}$), where 
			$\phi_{01}=\phi_{12}=0$, $\sigma \in [13.5, 45]$ ns and $\kappa \in [1.5, 3]$; the pulse sequence is truncated in the middle 
			of the Gaussian drive $\Omega_{01}(t)$ such that T$_{01}$=$3\sigma+\kappa \sigma$.			
			Results from the simulations are obtained (a1, b1) without decoherence, (a2, b2) with decoherence, and the phase $\phi_{02}$ is constant in time. Vertical dashed line in each plot corresponds to $\phi_{02}=\pi/2$.
	}
	\label{fig:phase}
	\label{fig:phase_with_time}
\end{figure}

We simulate saSTIRAP protocols for different phases of the counter diabatic drive ($\phi_{02}$), with $\phi_{01}=\phi_{12}=0$. For best transfer in a 
	saSTIRAP, $\phi_{01}+\phi_{12}+\phi_{20}=-\pi/2$~\cite{antti-2019} which is reflected in
	Fig.~\ref{fig:phase}(a1,a2) presenting the maximum population attained in the second excited state ($p_2$); given different values of the initial phases $\phi_{02}$ and standard deviations $\sigma$ of the Gaussian drives in the range $\sigma \in [13.5, 57.5]$ ns, without
	and with decoherence respectively. The value of $\kappa$ is fixed at $\kappa=2.7$ such that $t_s=-2.7 \sigma$. 

	In Fig.~\ref{fig:phase_with_time}(b1,b2) the vertical axis represents the time interval $T_{01}$ which extends from the starting of the saSTIRAP pulse sequence ($t=t_i$) to the middle of the $0-1$ drive. Here $\sigma \in [13.5, 45]$ns, $\kappa \in [3, 1.5]$, such that $T_{01}=n \sigma + \kappa \sigma$, where $n=3$ (the Gaussians are truncated at $\pm n \sigma $). The range of $T_{01}$ is taken along the diagonal connecting ($\sigma=13.5$ ns, $\kappa=3.0$) to ($\sigma=45$ ns, $\kappa=1.5$) in Fig.~\ref{fig:k_sigma}.
	Fig.~\ref{fig:phase_with_time}(b1,b2) show the maximum population in the second excited state without and with decoherence respectively, 
	with constant phases of the drives.
	This surface plot has a flattened profile for maximum population transfer around $\phi_{02}=\pi/2$. The distribution is more sensitive to $\phi_{02}$ around an optimal transfer time, $T_{01}=170$ ns, which also corresponds to maximal population transfer in STIRAP.

\section{Experimental results}

\subsection{STIRAP vs saSTIRAP}

An ideal STIRAP may be designed for a perfect transfer of the population from the ground state 
to the second excited state in a three-level system. On the other hand, sa-STIRAP is much more 
robust against the parameters such as amplitudes, standard-deviations, and the relative separation 
between the pair of counter-intuitive Gaussian profiles, wherein the imperfections of these parameters is 
accounted for by an additional two-photon resonance.

	
	\begin{figure}[ht]
		\centering
		\includegraphics[scale=1.1]{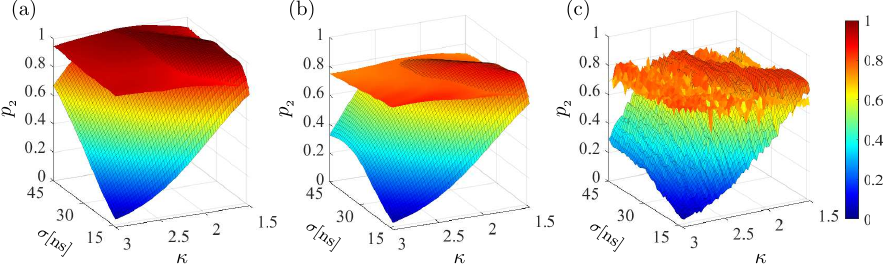}	
		\caption{Population $p_2$ in the state $\ket{2}$
				(a) ideally expected, (b) simulated with decoherence and (c) experimental data  for the STIRAP process and saSTIRAP with parameters $\Omega_{01}/(2\pi) = 44$ MHz, $\Omega_{12}/(2\pi) = 37$ MHz as a function of the pulse width $\sigma$ and the normalized pulse separation $\kappa = |t_{\rm s}|/\sigma$. Optimal populations resulting from STIRAP are shown with black mesh and for saSTIRAP without mesh.}
		\label{fig:k_sigma}
	\end{figure}
	

Next, we study experimentally the effect of changing the STIRAP pulse width $\sigma$ and the normalized STIRAP pulse separation $\kappa = |t_s|/\sigma$, see Fig. \ref{fig:k_sigma}.
For the data presented  we took $\Omega_{01} = 2\pi \times 44$ MHz and $\Omega_{12} = 2\pi \times 37$ MHz, which are both much smaller than the qubit anharmonicity $2\Delta = 2\pi \times 296$ MHz. 
Gaussian drives are truncated at $n=\pm3$.
It is clear from Fig.~\ref{fig:k_sigma} that saSTIRAP is insensitive to variations in $\sigma$ and $\kappa$ for a significantly wide range of values, whereas STIRAP typically works well when the pulses are relatively close to each other, corresponding to $|t_{\rm s}|/\sigma = 1.5$. Figs.~\ref{fig:k_sigma} (b,c) show that simulation and experimental results are in very good agreement with each other. These plots are obtained under strong decoherence, whose effect is partially mitigated by optimally truncating the STIRAP and saSTIRAP sequences in the middle of the drive $\Omega_{01}(t)$. By this time, the maximum transfer has already taken place and beyond this, the decay of $p_2$ dominates over the slow transfer of $p_0$ to $p_2$. Fig.~\ref{fig:k_sigma}(a) presents the ideal $p_2$ resulting from ideal STIRAP and saSTIRAP drives in the absence of decoherence, wherein there is a similar decline in STIRAP population transfer.

The plots which include the counterdiabatic correction demonstrate that the protocol is effective in counteracting the diabatic excitations, and we reach experimental values for $p_2$ in the range 0.8 - 0.9. These results can be also analyzed from the point of view of the quantum speed limit, by looking at the transfer time $t_{\rm tr}$ over which population transfer occurs. We define the transfer time as 
\begin{equation}
t_{\rm tr}= t_{f} - t_{i} = (2n+\kappa) \sigma \label{eq:quantum_speed}
\end{equation}
between an initial dark state $|D (t_{i})\rangle$ and a final state $|D (t_{f})\rangle$. A convenient choice for a dissipative system is to take an initial state with
99 \% population in $\ket{0}$ (mixing angles $\Theta (t_{i}) = 0.03 \pi$) and a final state with 80 \% population in $\ket{2}$ (mixing angle $\Theta(t_{f}) = 0.35 \pi$). For dissipationless systems the latter is usually taken 90\%, see \cite{Giannelli14}.

To further confirm these results, we realized another set of experiments where we observed the transfer efficiency of saSTIRAP against STIRAP area ($\mathcal{A}$) by varying the width ($\sigma$) of the STIRAP drive at constant amplitudes $\Omega_{01}/(2\pi) = 25$ MHz and $\Omega_{12}/(2\pi ) = 16$ MHz. We perform five different experiments where the aim was to reach a target value $p_2 = 0.55$ by adjusting the parameters ($\sigma$, $\mathcal{A}_{02}$). We found experimentally ($10$ ns, $\pi/2$), ($17$ ns, $3\pi/2$), ($17$ ns, $\pi/3$), ($25$ ns, $5\pi/4$), and ($25$ ns, $0$). The experimental points correspond well to the simulation.

\begin{figure}[ht]
	\begin{center}
		\includegraphics[width = 1\columnwidth]{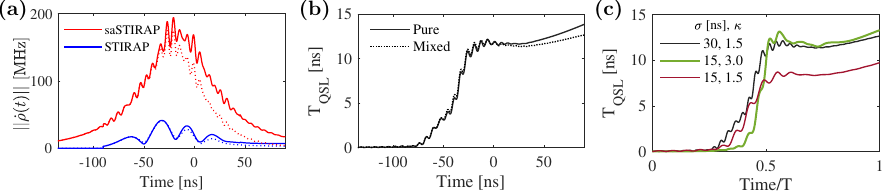}
	\end{center}
	\caption{(a) Instantaneous rate of change of quantum state resulting from  STIRAP and saSTIRAP evolutions without (continuous) and with (dotted) decoherence. Here $\sigma=30$ ns, $\kappa=1.5$, $\Omega_{01}/2\pi=44$ MHz, $\Omega_{12}/2\pi=37$ MHz, and $n=\pm 3$.  (b) Corresponding quantum speed limit without (continuous) and with (dotted) decoherence as a function of time, and (c) $T_{QSL}$ resulting from saSTIRAP with decoherence for different values of $\sigma$ and $\kappa$.} \label{Fig-qsl}
\end{figure}

Several bounds on the speed of state transfer have been derived in open systems, for example based on Fisher information \cite{Taddei}, on relative purity \cite{delCampo}, and on the Bures distance \cite{Lutz, Wu2020}.
	To find the quantum speed for evolution from a pure initial state $\rho_{0}=|\psi (0)\rangle \langle \psi (0)|$ at $t=0$ to a mixed state $\rho_{\tau}$ at $t=\tau$, we use the bounds derived in Ref. \cite{Lutz},
	\begin{equation}
	T_{\rm QSL}=\frac{\sin^2 (d_{B}(\rho_{0},\rho_{\tau}))}{\lambda_{\tau}^{\rm op}}
	\end{equation}
	where $d_{B}(\rho_{0},\rho_{\tau}) = \arccos (\sqrt{F(\rho_{\tau}, \rho_{0})})$ is the Bures distance and $F(\rho_{\tau}, \rho_{0}) = \left({\rm tr}\sqrt{\sqrt{\rho_{0}}\rho_{\tau}\sqrt{\rho_{0}}}\right)^{2}$ is the fidelity ($F(\rho_{\tau}, \rho_{0}) = \langle \psi (0)|\rho_{\tau} |\psi (0)\rangle $ for a pure $\rho_{0}$).
	This yields
	\begin{equation}
	T_{\rm QSL}=\frac{1-F(\rho_{\tau}, \rho_{0})}{\lambda_{\tau}^{\rm op}}
	\end{equation}
	and 
	\begin{equation}
	\lambda_{\tau }^{\rm op}= \frac{1}{\tau}\int_{0}^{\tau} ||\mathbb{L} [\rho (t) ]||_{\rm op}dt
	\end{equation}
	and $||\mathbb{L}[\rho (t) ]||_{\rm op}={\rm max}_{i}\{s_{i}(t)\}$ 
	is the operator norm ($s_{i}$ are the singular values of 
	$L [\rho (t)]$ defined as eigenvalues of $\sqrt{\mathbb{L}^{\dag} [\rho (t)]\mathbb{L} [\rho (t)]}$, where $\mathbb{L}^{\dag}$ is the adjoint of $\mathbb{L}$). 
	We obtain the Bures distance and corresponding quantum speed for the mixed state evolution under STIRAP and saSTIRAP drives (see Eqns.~\ref{eq:H01}, \ref{eq:H12}, \ref{eq:H2ph}), with $\sigma=30$ ns, $\kappa=1.5$, $\Omega_{01}/2\pi=44$ MHz, $\Omega_{12}/2\pi=37$ MHz, and $n=\pm 3$.
	In the ideal situation, where the initial state is $|0\rangle$ and the final state is $-|2\rangle$, the Bures distance between the initial and the final state is $0.5\pi$. However, with given STIRAP and saSTIRAP parameters, the maximum Bures distance is $0.48 \pi$ without considering decoherence and $0.33\pi$ in the presence of decoherence.
	Correspondingly, the instantaneous rate of change of quantum state also varies with and without decoherence as shown in Fig.~\ref{Fig-qsl}(a), where continuous curves are obtained by considering the pure state evolution (no decoherence) and dotted curves are obtained from the mixed state evolution (with decoherence). The rate of change $||\dot{\rho}(t)||$ of a quantum state  at an arbitrary time ($t$) is 
	defined as the 
	operator norm of $\dot{\rho}(t)$, which is in fact the same as $||\mathbb{L}[\rho (t) ]||_{\rm op}$, see Eq.~(\ref{eq:2.5}). It is noteworthy that the rate of change of quantum state in saSTIRAP (in red) is much higher than that of STIRAP (in blue), while both of these peak around the middle of the sequence. In Fig.~\ref{Fig-qsl}(b), the quantum speed limit ($T_{QSL}$) under saSTIRAP drive is shown as a function of time for pure and mixed states. In case of pure state evolution, $T_{QSL}$ reaches a maximum of $13.9$ ns while the mixed state evolution has a lower bound at $T_{QSL}=12.6$ ns. This faster mixed state evolution is deceptive, as it results from the smaller Bures distance and infact has lower value of $||\dot{\rho}(t)||$. 

	Considering Fig.~\ref{Fig-qsl}(a) again, it is evident that $T_{QSL}$ for STIRAP is an order of magnitude larger than that of saSTIRAP, which is attributed to almost the same Bures distance while an order of magnitude difference exists in the rate of evolution of the quantum states.  Finally, we compare the quantum speed limits of the mixed state evolutions for different STIRAP parameters: ($\sigma=15$ ns, $\kappa=3$), where saSTIRAP evolution is predominantly achieved by the counterdiabatic drive, as can be seen from Fig.~\ref{fig:k_sigma} and 
	for ($\sigma=15$ ns, $\kappa=1.5$), which corresponds to much shorter pulse duration. Results from the comparisons are shown in Fig.~\ref{Fig-qsl}(c), where red curve presents the fastest evolution with $T_{QSL}=9.6$ ns.
	Another interesting point would be to compare the quantum state evolution under STIRAP and saSTIRAP for ($\sigma=15$ ns, $\kappa=3$), where ill performance of STIRAP is readily seen in Fig.~\ref{fig:k_sigma}(a) as well as from $d_B(\rho_{t_i}, \rho_{t_f})=0.19 \pi$ (without decoherence). In this case, the maximum value of the rate of change of quantum state under STIRAP ($||\dot{\rho}(t)|| \approx 215$ MHz) is of the same order of magnitude as that of saSTIRAP ($||\dot{\rho}(t)|| \approx 250$ MHz). However, in the same time $T=t_f-t_i$, STIRAP leads to nowhere close to the expected final state, while saSTIRAP driven $\rho_{t_i}$ is close to the final state with $d_B(\rho_{t_i}, \rho_{t_f})=0.43 \pi$.

For pure states the Bures distance becomes the Fubini-Study distance
	\begin{equation}
	d_{\mathrm{FS}}(|\psi_{1}\rangle, 
	|\psi_{2}\rangle) = \cos^{-1} \sqrt{F(|\psi_{1}\rangle, 
		|\psi_{2}\rangle)}
	\end{equation}
	where the fidelity is defined as $F(|\psi_{1}\rangle, 
	|\psi_{2}\rangle) = |\langle \psi_{1}|\psi_{2}\rangle|$. 
	In the case of dark states $|D\rangle = \cos \theta |0\rangle - \sin \Theta |1\rangle$ we have 
	\begin{equation}
	d_{\mathbf{FS}} (\Theta_{1}, \Theta_{2}) = |\Theta_{1}-\Theta_{2}|^{2}
	\end{equation}
	and the Bures distance is $d_{\mathbf{Bures}} = \sqrt{2 (1 - \cos (\Theta_{1}-\Theta_{2}))}$.

\subsection{Pulse area}

\begin{figure}[htb]
	\centering
	\includegraphics[width = 0.8\textwidth]{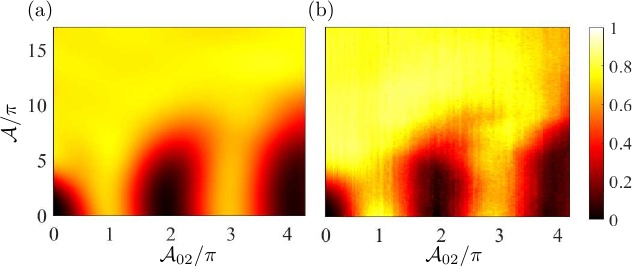}
	\caption{
	Transferred population $p_2$ from 
	a) simulation, b) experiment- 
	as a function of the STIRAP pulse area $\mathcal{A}$ and the two-photon pulse area $\mathcal{A}_{02}$.
	The STIRAP Rabi couplings are varied linearly from zero to a maximum value of $\Omega_{01}/(2\pi) = 80$ MHz and $\Omega_{12}/(2\pi)= 58$ MHz while keeping the asymmetry constant at $\eta = 0.16$. Similarly, the two-photon pulse was varied from zero to maximum values of $\tilde{\Omega}_{12}/(2\pi )= \sqrt{2}\tilde{\Omega}_{01}/(2\pi ) = 118$ MHz. The separation time was $t_{\rm s} = -45$ ns and the width was $\sigma = 30$ ns. 
	}
	\label{fig:robustness}
\end{figure}

STIRAP is known to be insensitive to the amplitudes of the drives once it satisfies the adiabaticity criteria, $\Omega \sigma >> \sqrt{\pi}/4$ (assuming $\Omega_{01}=\Omega_{12}=\Omega$). Thus for a fixed $\sigma$, there is a threshold $\Omega$ beyond which the value of $\Omega$ ceases to matter anymore in a STIRAP. This robustness feature of STIRAP is passed onto saSTIRAP, as seen in  Fig.~\ref{fig:robustness}, where the population transferred from the ground state to the second excited state $p_2$ is plotted as a function of the STIRAP area, $\mathcal{A}=\int_{-\infty}^{\infty} \sqrt{\Omega_{01}(t)^2 +  \Omega_{12}(t)^2 } dt$ and two-photon pulse area, $\mathcal{A}_{02}=2\Theta (t_f)$.
The STIRAP area is spanned by linearly varying the drive-amplitudes in the ranges: $\Omega_{01} \in 2\pi \times [0, 80]$ MHz and $\Omega_{12} \in 2\pi \times [0, 58]$ MHz with asymmetry $\eta=(\Omega_{01}-\Omega_{12})/( \Omega_{12}+\Omega_{01} )=0.16$, while keeping $\sigma=30$ ns, $t_s=-1.5 \sigma$, $t_i=-3\sigma+t_s$, and $t_f=3\sigma$ fixed. 
The area of the two-photon drive is spanned by simply varying the target mixing angle $\Theta (t_f)$, such that $\mathcal{A}_{02} \in [0, 4\pi]$. In Fig.~\ref{fig:robustness}, $\mathcal{A}_{02}=0$ corresponds to purely STIRAP implementation. In that case, population transfer efficiency becomes significant for $\mathcal{A}>4\pi$. From the adiabaticity criteria, we have $\Omega>>14.8$ MHz and to an approximation $\mathcal{A}>>3.17\pi$. Fixing the STIRAP area $\mathcal{A}=4\pi$, and amplitude ratio, $a=(1-\eta)/(1+\eta)=\Omega_{12}/\Omega_{01}=0.725$, we obtain $\Omega_{01}/2\pi=18.6$ MHz and $\Omega_{12}/2\pi=13.48$ MHz. The efficiency of STIRAP and its robustness against the drive-amplitudes can be observed along the vertical stretch for $\mathcal{A}_{02}=0$. 
  The area of the counter-diabatic saSTIRAP drive,
  $\mathcal{A}_{02}=\int_{-\infty}^{\infty} \Omega_{02}(t) dt = \int_{t_i}^{t_f} 2\dot{\Theta}(t) dt = 2\theta(t_f)$, is equal to $\pi$ for saSTIRAP. 
 The drive $\Omega_{02}(t)$ has a maximum amplitude of $-t_s/\sigma^2$ at $t=t_s/2$, see Eq. (\ref{eq:omega02t}) and the corresponding amplitudes of two-photon drives can be obtained by using: 
  $\tilde{\Omega}_{12}=\sqrt{2}\tilde{\Omega}_{01}=\sqrt{\sqrt{2} \Delta \Omega_{02}}$. In the experiments and simulations a linear variation of $\tilde{\Omega}_{12} \in 2\pi [0,118]$ MHz corresponds to $\mathcal{A}_{02} \in [0, 4\pi]$.
As $\mathcal{A}_{02}$ is increased from zero, the population transfer begins to improve and for $\mathcal{A}_{02}=\pi$, we have saSTIRAP with a near perfect population transfer. This narrow region around $\mathcal{A}_{02}=\pi$ is clearly insensitive to the STIRAP area $\mathcal{A}$. As STIRAP begins to work properly, this narrow region gets wider and we obtain a highly efficient population transfer. With $A_{02}$ approaching $2\pi$, the counter-diabatic two-photon drive of saSTIRAP effectively corresponds to a $2\pi$ rotation in the $|0\rangle-|2\rangle$ subspace. For $\mathcal{A}_{02}=3\pi$, a saSTIRAP like transfer is observed again. Here the two-photon drive effectively implements a $3\pi$ rotation such that initial state $|0\rangle$ ends up close to $i|2\rangle$, which interferes constructively with the output state ($-|2\rangle$) from STIRAP.  
The phase difference between these two output states prevents them to interfere destructively. The periodicity of the pattern for relatively small values of $\mathcal{A}$ in Fig.~\ref{fig:robustness} is infact a consequence of Rabi oscillations in the $|0\rangle-|2\rangle$ subspace due to two-photon drives: $\tilde{\Omega}_{01}(t)$, detuned by $-\Delta$ from $\omega_{01}$ and $\tilde{\Omega}_{12}(t)$, detuned by $\Delta$ from $\omega_{12}$.

To summarize, we have demonstrated robustness under scaling for the superadiabatic protocol under typical experimental constraints related to noise and maximally-achievable drive amplitudes.

\section{Discussion: comparison with saSTIRAP by direct-coupling counterdiabatic drive}

\begin{figure}[htb]
	\centering
	\includegraphics[width = 0.9\textwidth]{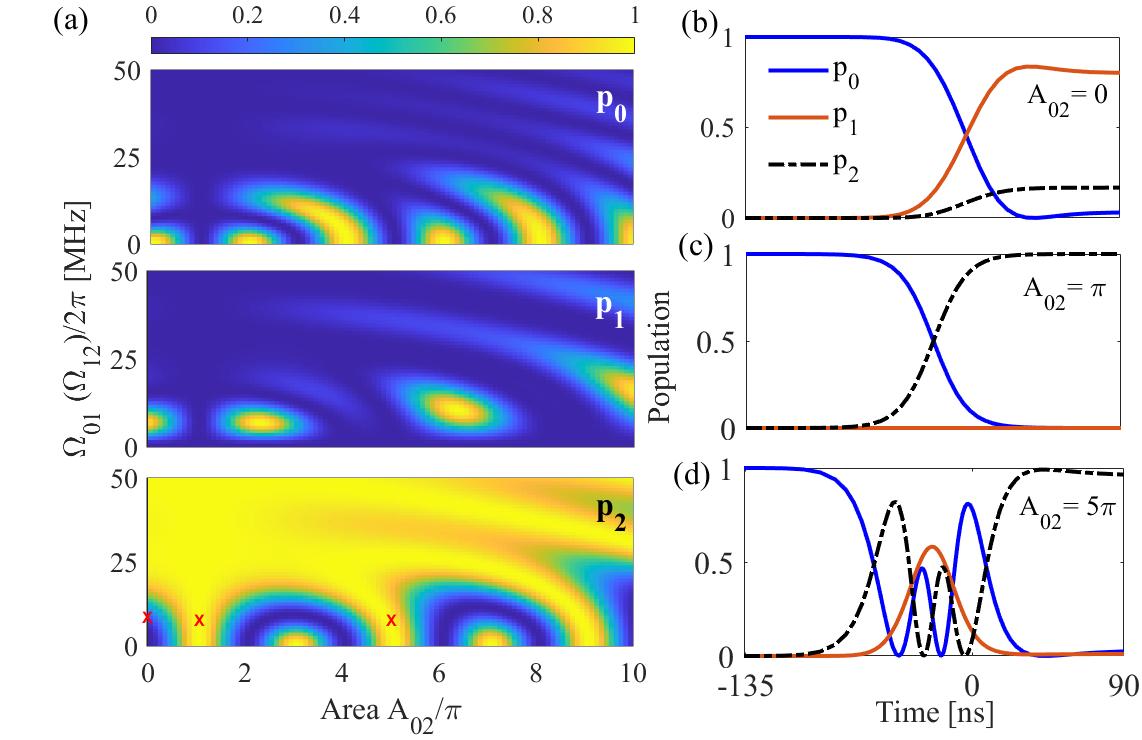}
	\caption{Results from simulations of saSTIRAP with a direct $0-2$ counterdibatic drive and in the absence of decoherence. (a) Final populations - $p_0$, $p_1$, and $p_2$ resulting from saSTIRAP, as a function of amplitudes of STIRAP drives ($\Omega_{01}$, $\Omega_{12}$) and area of the superadiabatic drive ($\mathcal{A}_{02}$). Here $\sigma=30$ ns, $t_s=-45$ ns, $t_i=-3\sigma+t_s$, $t_f=3\sigma$, $\phi_{01}=\phi_{12}=0$, and $\phi_{20}=-\pi/2$. Corresponding population-transfer profiles over time are plotted for $\Omega_{01}=\Omega_{12}=2\pi \times 8$ MHz with (b) $\mathcal{A}_{02}=0$, (c) $\mathcal{A}_{02}=\pi$, and (d) $\mathcal{A}_{02}=5\pi$. These points are also marked with red cross marks in the surface map of $p_2$.
	}
	\label{fig:robustness_direct}
\end{figure}

In order to understand better the previous results, we analyze here the case of saSTIRAP using a counterdiabatic drive that couples directly with transition $|0\rangle-|2\rangle$.
This is the standard way to obtain a highly efficient desired population transfer. We study here what happens when the strength of this counterdibatic drive area exceeds the value of $\pi$ obtained in the standard superadiabatic theory.

The superadiabatic(sa)-STIRAP closely follows the dynamics of adiabatic population transfer via STIRAP while correcting for any non-adiabatic excitations.
The counterdiabatic ($0-2$) drive can be understood as working in parallel with the STIRAP sequence; they interfere constructively to achieve the final 
dark state $|\psi_f\rangle = |D\rangle= \cos (\Theta(t_f)) |0\rangle - \sin (\Theta(t_f)) |2\rangle$ with mixing angle $\Theta (t_f)=2n\pi+\pi/2$, where $n \in \mathbb{Z}$. 
 saSTIRAP is strictly defined for $\mathcal{A}_{02}=\pi$ and $\phi_{20}=-\pi/2$ where the phases $\phi_{01}$ and $\phi_{12}$ of the STIRAP pulses are fixed to $0$ in Eq. (\ref{eq:H01}) and Eq. (\ref{eq:H12}). 
However, now the 0-2 drive, which in Eq. (\ref{eq:H02}) was obtained from the effective two-photon drive, is now realized directly as $H_{02}  = (i\hbar \Omega_{02}/2)\left(|0\rangle \langle 2| - |2\rangle \langle 0|\right)$. Importantly, this means that energy-level shifts, which inevitably appear in the two-photon drive, do not appear. We simulate the dynamics of a three-level system initialized in the dark state
$|\psi_i\rangle = |D\rangle= \cos (\Theta(t_i)) |0\rangle - \sin (\Theta(t_i)) |2\rangle$ with mixing angle $\Theta (t_i)=0$. The decoherence is neglected altogether. We extend the simulation to a much broader range of $\mathcal{A}_{02}$ beyond the standard area of $\mathcal{A}_{02}=\pi$ and observe a periodically repeating pattern with features similar to saSTIRAP, see Fig. \ref{fig:robustness_direct}.

The counterdiabatic pulse area 
$\mathcal{A}_{02}= \int_{-\infty}^{\infty} \Omega_{02}(t) dt = 2\int_{-t_i}^{t_f} \dot{\Theta}(t) dt = 2\Theta(t_f)$ since $\Theta(t_i)=0$. Therefore, the final angle is $\Theta(t_f)=\mathcal{A}_{02}/2$ and the corresponding final state is
\begin{equation}
	|\psi_f(\mathcal{A}_{02})\rangle = \cos (\mathcal{A}_{02}/2) |0\rangle + e^{i\phi}  \sin (\mathcal{A}_{02}/2) |2\rangle \label{eq:finalstate}
\end{equation}
which is a dark state ($|D(\theta)\rangle=\cos \theta |0\rangle - \sin \theta |2\rangle$) for $\mathcal{A}_{02}=2n\pi+\theta$ and $\phi=(2m\pi+\phi_{20}-\pi/2)$, where $n,m \in \mathbb{Z}$.
Populations of the 
ground state -- $p_0$, first excited state -- $p_1$, and second excited state -- $p_2$ obtained in a saSTIRAP are plotted with resect to maximum amplitudes in the STIRAP drive ($\Omega_{01}=\Omega_{12}$) and area of the counterdiabatic pulse- $\mathcal{A}_{02}$, as shown in Fig.~\ref{fig:robustness_direct}. Simulations in Fig.~\ref{fig:robustness_direct} have STIRAP drives with fixed width $\sigma=30$ ns  and overlap $t_s=-1.5 \times \sigma$ with each Gaussian being truncated at $\pm 3\sigma$.
 In Fig.~\ref{fig:robustness_direct}(a), $\Omega_{01}=0$ corresponds to Rabi oscillations in the $0-2$ subspace, such that $p_0+p_2=1$, while $p_1=0$. $\mathcal{A}_{02}=0$ corresponds to the STIRAP implementation alone, which begins to work well beyond $\Omega_{01}=2\pi \times 15$ MHz. STIRAP population dynamics with time, in the region of inefficient implementation with $\Omega_{01}=\Omega_{12}=2\pi \times 8$ MHz is shown in Fig.~\ref{fig:robustness_direct}(b). As the counterdiabatic pulse of saSTIRAP comes into picture, there is a dramatic improvement with a perfect population transfer as shown in Fig.~\ref{fig:robustness_direct}(c). 
 Interestingly, highly efficient population transfer appears also at other values of $\mathcal{A}_{02}$, most notable near values $\mathcal{A}_{02}=4n\pi + \pi$ ($n \in \mathbb{Z}$, $n\neq 0$). An example of time-domain population transfer is shown in Fig.~\ref{fig:robustness_direct}(d) for $\mathcal{A}_{02}=5\pi$. However, it is important to notice that in this case has additional oscillations in the middle of the sequence, therefore the non-adiabatic terms are not suppressed. Still, such protocols are also interesting and can be used to realize holonomic gates, see for example Ref. \cite{Danilin2018} for an experimental realization.

 Another interesting situation arises for $\mathcal{A}_{02}=3\pi$, where there is destructive interference between states resulting from STIRAP and counter-diabatic drive for $\Omega_{01}=\Omega_{12}=2\pi \times 8$ MHz. The destructive interference is  due to an overall phase of $e^{i\pi}$ acquired by the target dark state in the case of direct 0 -- 2 drive (with $\mathcal{A}_{02}=3\pi$), \emph{i.e.} under $H_{02}(t)$ drive the corresponding final states, see Eq.~(\ref{eq:finalstate}) for $\mathcal{A}_{02}=\pi$ and $3\pi$ are related as
 $|\psi_f(\pi) \rangle = e^{i\pi}|\psi_f(3\pi)\rangle$.

 This interference is absent in the results discussed in previous section with reference to Fig.~\ref{fig:robustness}. In fact under two-photon drive the state does not develop an overall phase of $e^{i\pi}$ for the case of $\mathcal{A}_{02}=\pi$ vs $\mathcal{A}_{02}=3\pi$.

 The final state from the two-photon drive contains both real and imaginary parts, which do not get nullified by the final state resulting from STIRAP, which has only real coefficients. 
 The difference between the actions of direct 0 -- 2 drive and the two-photon drive can be explained by considering the details of the effective two-photon drive in the $0-2$ subspace. Let us consider the evolution of an arbitrary state $\alpha |0\rangle + \beta |1\rangle + \gamma |2\rangle$ driven by two-photon drive Hamiltonian given in Eq.~(\ref{eq:H2ph}). Here we use the method of adiabatic elimination by assuming $\Omega_{02}<<\Delta$ and $\dot{\beta}=0$ as described in Ref.~\cite{Vepsalainen2018}. This leads to an effective Hamiltonian in the $0-2$ subspace,
 \begin{eqnarray}
 	H_{\rm 2ph}^{\rm eff}(t) &=& -\frac{\hbar \tilde{\Omega}_{01}\tilde{\Omega}_{12}}{4 \Delta} \left( \frac{1}{2\lambda} |0\rangle \langle 0| + 
 	\frac{\lambda}{2} |2\rangle \langle 2| + e^{2i\tilde{\varphi}}|0\rangle \langle 2| \right)  + h.c..
 	\label{eq:H2pheff}
 \end{eqnarray}
This Hamiltonian is clearly different from the directly coupled drive $H_{02}(t)$ from Eq.~(\ref{eq:H02} in the $0-2$ subspace. These additional diagonal terms in Eq.~(\ref{eq:H2pheff}) are responsible for generating additional relative phases during the evolution, which lead to different phases of the complex coefficients, even though their absolute values remain close to the ones expected from a direct 0--2 drive.

 In addition, from Fig.~\ref{fig:robustness}(a) and Fig. \ref{fig:robustness_direct}(a), imperfections in the $p_2$ surface maps in the form of tilted arcs with tails for increasing values on $\mathcal{A}_{02}$ occur. This happens because $\Omega_{02}$ becomes comparable with the anharmonicity $\Delta$, and also because of higher truncation errors due to increasing $0-2$ pulse amplitude. Despite these imperfections, a much wider plateau of efficient population transfer is formed, featuring insensitivity to pulse parameters. The structure of this plateau is preserved even in the presence of decoherence. An added advantage is seen for $\mathcal{A}_{02}=\pi$, $5\pi$ in Fig.~\ref{fig:robustness_direct}(a), and $A_{02}=\pi$, $3\pi$ in Fig.~\ref{fig:robustness}, where population transfer is resilient to the STIRAP drive amplitudes. 
 We have verified that a similar plateau structure of efficient population transfer can be obtained for constant $\Omega_{01}$ and $\Omega_{12}$ by varying the total pulse duration, which is a combination of $\sigma$, $\kappa$, and $n$. The plateau of efficient population transfer in this case is resilient to errors in the total pulse duration.
 The large flexilibility in the choice of parameters without compromising the efficiency of the protocol makes the protocol remarkably robust.


\section{Conclusion}

In the loop conﬁguration for a transmon device, we have implemented a superadiabatic protocol where two couplings produce the standard stimulated Raman adiabatic passage, while the third is a counterdiabatic ﬁeld that suppresses the nonadiabatic excitations. 
This technique enables fast operation and it is remarkably robust against errors in the shape of the pulses.
 The superadiabatic method enables a continuos interpolation between speed and insensitivity to errors, allowing one to select the optimal values under realistic experimental constraints. We also observe the appearance of plateaus characterized by highly efficient transfer of population at large values of the counterdiabatic pulse area.

\vskip6pt
\enlargethispage{20pt}


\vspace{20mm}

{\bf Data accessibility} Data and codes used in this study can be accessed from the GitHub repository available at https://github.com/dograshruti/Robustness-of-saSTIRAP.

{\bf Author's contributions} S.D.: conceptualization, data curation, formal analysis, investigation, methodology, software, visualization, writing—original draft, writing—review and editing; A.V.: data curation, formal analysis, investigation, methodology, software; G.S.P.: conceptualization, funding acquisition, methodology, project administration, supervision, visualization, writing—original draft, writing—review and editing. All authors gave final approval for publication and agreed to be held accountable for the work performed therein.

{\bf Conflict of interest declaration} The authors declare that they have no competing interests.

{\bf Funding} We acknowledge financial support from the Finnish Center of Excellence in Quantum Technology QTF (projects 312296, 336810) of the Academy of Finland. We also are grateful for financial support from the RADDESS programme (project 328193) of the Academy of Finland and from Grant No. FQXi-IAF19-06 (“Exploring the fundamental limits set by thermodynamics in the quantum regime”) of the Foundational Questions Institute Fund (FQXi), a donor advised fund of the Silicon Valley Community Foundation. This project has received funding from the European Union’s Horizon 2020 research and innovation programme under grant agreement no. 824109 (European Microkelvin Platform project, EMP). This work used the experimental facilities of the Low Temperature Laboratory of OtaNano.

{\bf Acknowledgments} We thank Dr. Sergey Danilin for fabricating the sample used in these experiments and for assistance with the measurements.

\vskip6pt

\appendix

\section{Appendix: Dynamic phase corrections}

\begin{figure}[h]
	\centering
	\includegraphics[scale=1.1,keepaspectratio=true]{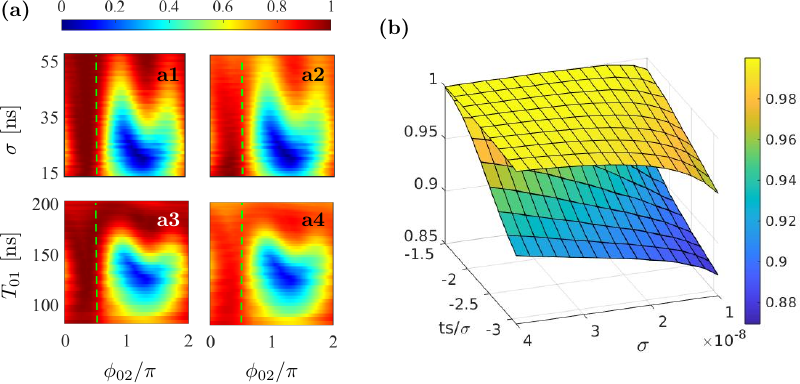}
	\caption{(a) Population $p_2$ resulting from saSTIRAP with dynamically varying phases of the drives is plotted as a function of initial phase of the $0-2$ drive. (a1, a3), (a2, a4) show the transferred population ($p_2$) without and with decoherence respectively in case of 
		dynamically varying phase ($\phi_{02}(t)$). Vertical dashed line in each plot corresponds to $\phi_{02}=\pi/2$. $p_2$ as a function of $\kappa$ and $\sigma$ in the case of saSTIRAP with constant phases of the drives and with dynamic phases of the drives. (b) $p_2$ from saSTIRAP driven dynamics as a function of $\kappa$ and $\sigma$.
} \label{fig:app2}
\end{figure}

	Following Ref. ~\cite{Vepsalainen2018}, dynamic phase corrections implemented to drives $\Omega_{01}(t)$, $\Omega_{12}(t)$, and $\Omega_{02}(t)$ (with $\mathcal{A}_{02}=\pi$) are 
	$\phi_{01}(t) = \phi_{01} + 2\sqrt{2} \hbar \Theta(t)$, $\phi_{12}(t) = \phi_{12} - (5 \hbar/\sqrt{2}\Theta(t)$, and $\phi_{02}(t) = \phi_{02} - ( \hbar/\sqrt{2}\Theta(t)$ respectively. 
	We simulated saSTIRAP with dynamically varying phases by incorporating these ac-Stark shift corrections.
	Fig. ~\ref{fig:app2}(a) presents the resulting surface maps of $p_2$ as a function of initial phase of the $0-2$ drive plotted with $\sigma$ in parts (a1,a2) and as a function of the transfer time in parts (a3, a4), similar to that of Fig.~\ref{fig:phase}.
	Interestingly, by attributing time-dependent dynamical phases to the drives in order to compensate the ac-Stark shifts resulting from the two-photon resonanant drive, intial choice of $\phi_{02} (t=0)=\pi/2$ is not necessarily optimal. A clear shift in the initial $\phi_{02}(t)$ is observed as shown in Fig.~\ref{fig:phase}(a3,a4) without and with decoherence.

	In Fig.~\ref{fig:app2}(b), results from saSTIRAP driven dynamics is shown as 3D map of $p_2$ as a function of $\kappa$ and $\sigma$. The map in which $p_2$ decreases for smaller values of $\sigma$ and relatively larger $t_s/\sigma^2$ is acquired with constant phases $\phi_{01}=\phi_{12}=0, \phi_{20}=-\pi/2$. The map acquired with dynamic phase corrections as described above has higher values for the shown range.
	Thus, ac-Stark shift compensations using dynamic phases of the drives results into highly efficient population transfer.
	Constant phases of the drives are at their worst performance for small $\sigma$ values, which is the region where STIRAP drives are also not effective. In this region, the saSTIRAP efficacy highly depends upon the $0-2$ drive. Therefore, this region faces maximum errors due to the detuned two-photon drives and hence there is a lot of room for the improvement by compensating the corresponding ac-Stark shifts.

\section{Appendix: Majorana representation}
\begin{figure}
\centering
\includegraphics[scale = 1.2]{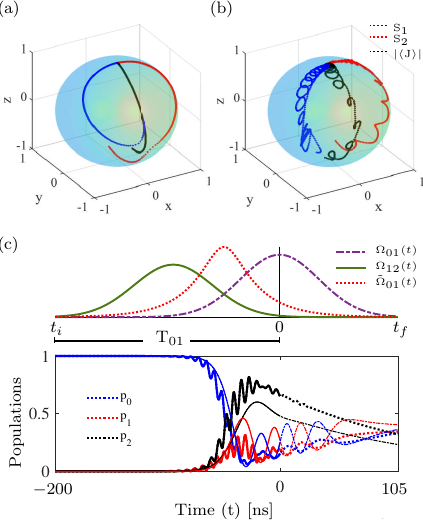}   
\caption{Majorana trajectories of a qutrit driven by (a) STIRAP, (b) saSTIRAP are shown, where the red and blue trajectories follow two Majorana stars representing a qutrit and the black trajectory is the path traversed by the averaged magnetization vector. 
(c) saSTIRAP drives and corresponding variation of populations in levels $|0\rangle$ (black), $|1\rangle$ ( red), and $|2\rangle$ (blue) with optimal transfer time and total transfer time are shown with solid lines (thick for saSTIRAP and thin for STIRAP) and dashed lines (for STIRAP or dotted for saSTIRAP)   respectively. 
}
\label{fig:Majorana}
\end{figure}

To compare further the efficacies of STIRAP and saSTIRAP, we present a geometrical visualization of the results from STIRAP and saSTIRAP on the Majorana sphere~\cite{dogra-2020}. The transmon, initialized in its ground state ($|0\rangle$), is driven by the STIRAP and saSTIRAP Hamiltonians with $\sigma=35$ ns, $\kappa=2.7$ and $n=\pm 3$ in a decohering environment. Subject to the strong decoherence (as mentioned in the main text), the mixedness keeps on increasing and the purity of the single qutrit state drops to as low as 0.45. However, by considering a optimal truncation of the pulse sequence, which falls close to the middle of the $0-1$ Gaussian drive, the purity saturates close to $0.8$. 
Majorana trajectories in Fig.~\ref{fig:Majorana}(a), (b) present the mixed state dynamics for the maximum population transfer, for STIRAP and saSTIRAP drives respectively. Here we show the trajectories followed by the dominant eigenstates alone, as other two eigenvalues are order of magnitude lower. 
The corresponding variation of populations in levels $|0\rangle$ (black lines), $|1\rangle$  (red lines), and $|2\rangle$ (blue lines) are shown in Fig.~\ref{fig:Majorana}(c). The optimally truncated dynamics is shown with continuous lines (thick for saSTIRAP and thin for STIRAP) until a maximum transfer of population from $|0\rangle-|2\rangle$ is achieved, while the respective curves continue as dashed (STIRAP) and dotted (saSTIRAP) lines. The levels of decoherence and other
experimental parameters of STIRAP and saSTIRAP are the same, but the results from saSTIRAP are clearly better.

\end{document}